\documentclass[fleqn,10pt]{wlscirep}

\usepackage{xr}
\usepackage{cleveref}
\usepackage{soul,xcolor}
\usepackage{amsmath}

\setstcolor{red}

\title{Conformal printing of graphene for single and multi-layered devices on to arbitrarily shaped 3D surfaces}
\author[1]{Leonard W. T. Ng}
\author[1]{Xiaoxi Zhu}
\author[1]{Guohua Hu}
\author[1]{Nasiruddin Macadam}
\author[1]{Dooseung Um}
\author[1]{Tien-Chun Wu}
\author[2]{Frederic Le Moal}
\author[2]{Chris Jones}
\author[1, *]{Tawfique Hasan}
\affil[1]{Cambridge Graphene Centre, University of Cambridge, Cambridge CB3 0FA, UK}
\affil[2]{Novalia Ltd, Impington, Cambridge CB24 9NP, UK}
\affil[*]{th270@cam.ac.uk}

\begin{abstract}
	Printing has drawn a lot of attention as a means of low per-unit cost and high throughput patterning of graphene inks for scaled-up thin-form factor device manufacturing. 
	However, traditional printing processes require a flat surface and are incapable of achieving patterning on to 3D objects. 
	Here, we present a conformal printing method to achieve functional graphene-based patterns on to arbitrarily-shaped surfaces. 
	Using experimental design, we formulate a water-insoluble graphene ink with optimum conductivity.
	We then print single and multi-layered electrically functional structures on to a sacrificial layer using conventional screen printing. 
	The print is then floated on water, allowing the dissolution of the sacrificial layer, while retaining the functional patterns. 
	The single and multilayer patterns can then be directly transferred on to arbitrarily-shaped 3D objects without requiring any post deposition processing. 
	Using this technique, we demonstrate conformal printing of single and multilayer functional devices that include joule heaters, resistive deformation sensors and proximity sensors on hard, flexible and soft substrates, such as glass, latex, thermoplastics, textiles, and even candies and marshmallows.
	Our simple strategy offers great promises to add new device and sensing functionalities to previously inert 3D surfaces. 
\end{abstract}

\begin{document}
	\flushbottom
	\maketitle
	\thispagestyle{empty}
	
	\section{Introduction}
	
	Although a number of methods have been explored for the deposition of 2D materials for functional devices, printing holds specific promise for high-volume, low-cost and large-area manufacturing \cite{Cui2016, Ng2019, Hu2018, Torrisi2012}.
	These include inkjet \cite{Hu2018, Torrisi2012, Mcmanus2017, Howe2015}, screen \cite{Joseph2016, Secor2014} and roll-to-roll (R2R) gravure and flexographic printing which are already widely used in the large-scale, ultra-low-cost production of packaging materials, everyday documents, magazines and newspapers \cite{Ng2019a}. 
	Among these, screen printing has emerged as an effective method in depositing high-viscosity graphene inks directly on to the substrates, typically \emph{via} a stencil made of a fine synthetic fibre mesh that is capable of depositing $\sim$10-16 $\mu$m thick layers of ink \cite{R.H.Leach1993, Ng2019a}. 
	Recent investigations have shown customised screen printing can achieve high resolution ($<\sim$40 $\mu$m) patterns of 2D materials on a variety of substrates, including glass, paper, textiles, polyester film and even solid graphite \cite{Zhang2017, Rowley-Neale2017, Hyun2015, Arapov2016, Xu2013, Yeates2017}.
	However, it is usually capable of printing only on to flat surfaces \cite{Hu2018, Ng2019a}.
	
	Conformal printing is an attractive and emerging field, which can enable specific electronic features such as sensors and circuits directly on to 3D objects.
	This is especially important in the present-day internet of things (IoT) era where 3D objects are increasingly interconnected.
	However, functional ink deposition on 3D structures is not a trivial process due to uneven and sometimes, complex topography.
	To date, several techniques have been explored to achieve conformal functional printing of 2D material, carbon nanotube and metal nanoparticle-based inks.
	These include aerosol jetting \cite{Jabari2015} or specially adapted inkjet or screen printing that manipulate arbitrarily-shaped substrates \emph{via} rotary systems around ink nozzle heads or printing mesh\cite{Krebs2009}. 
	However, these usually require highly complex and expensive printer designs that are capable of only depositing functional patterns on to one object at a time.
	Thermoforming has also been used in conjunction with other printing techniques such as flexography and gravure where inks are first deposited on planar polymer substrates which are then made pliable \emph{via} the application of heat and formed into a complex 3D shape in a multi-stage, offline process. \cite{Krebs2009}.
	Other more esoteric methods such as direct-write deposition \cite{Ahn2009} have also been explored.
	However, these demonstrations have been conducted only on a small scale, are capable of depositing only one material at a time and are highly time consuming.

We note that the general concept we present here has been previously demonstrated for the deposition or transfer printing of nanomaterials and functional structures on to flat and 3D objects.
In 2008, transfer deposition of 0D and 1D nanostructures \cite{Jiao2008} as well as atomically thin sheets of 2D materials \cite{Reina2008} on to flat substrates were demonstrated using poly(methyl methacrylate) (PMMA) as the sacrificial layer (SF).
Subsequently, this strategy was adapted for printing of nanostructures on to 3D objects using PMMA \cite{Lee2011} and other SFs such as polyvinyl alcohol (PVA) \cite{Aziz2011} and silk-fibroin \cite{Kim2010}.
Very recently, Saada \emph{et al.} used inkjet printed commercial silver inks to demonstrate patterns on to 3D objects by transfer printing using a PVA film as the SF \cite{Saada2017}.
However, this nanoparticle-based ink requires sintering to provide the desired functionality and mechanical integrity for the transfer process, potentially restricting the technique to single layer patterns only.

Among the various strategies discussed above, we propose that water or solvent-assisted printing is potentially the most promising method for large-area deposition of functional 2D material inks directly on to arbitrarily shaped objects.
This is due to the ability of this technique to conform to the complex topographies of 3D objects.
To date, such a method has not yet been developed for graphene and other 2D materials that is robust for large-area, single and multilayer devices for electronics and sensing.
	
Here, we introduce a method of conformal printing of a specifically formulated conductive graphene ink, exploiting the advantages of water-assisted conformal printing.
This method first involves the formulation of a water-insoluble ink that can be printed \emph{via} screen printing.
	For this, we use a combination of One Factor At A Time (OFAT) experiments (where we vary one factor in ink formulation and measure the corresponding outcome) and Design of Experiments (DOE), a method that we use to determine the inter-relationship between multiple components of the final ink formulation.
	OFAT experiments involve increasing the graphene loading until the ink is unprintable ($>$ 10 Pa s), while determining the minimum amount of binder required for the ink to be shear thinning during printing and to maintain its mechanical integrity on the substrate after curing.
	The results of OFAT are used to qualitatively define sensible upper and lower bounds for the DOE, which is then used to quantitatively optimise the electrical conductivity of the ink after printing and curing. 
	
	After formulation, the ink (and additional layers for the case of multi-layered devices) is printed on an ultrathin, water-soluble PVA as the SF. 
	The composite structure, comprising the SF and printed layer(s) is then floated in a water bath to allow the SF to dissolve. 
	The floating patterns are then `fished' out from the water-bath using the target substrates, allowing the pattern to adhere directly on to the substrate.
	This process is highly versatile, capable of printing on to different substrates to fabricate functional devices without the loss of the conductive properties of the printed graphene patterns after the transfer.
	
	We demonstrate this by conformally printing conductive graphene patterns to fabricate single and multi-layered devices on to multiple hard (glass, 3D printed thermoplastics), flexible and porous (bandages) and non-porous (nitrile gloves), and even on soft substrates such as marshmallows and gummy candies. 
	These structures are then used to demonstrate applications of conformal printing, including joule heaters on glassware, resistive sensors on off-the-shelf candies and electrodes on non-conductive objects, making them capable of capacitive proximity sensing. 
	We further demonstrate the capability of this method in the fabrication of multi-layered devices by printing graphene/polyurethane(PU)/graphene parallel-plate capacitors on arbitrarily shaped objects.
	
	\section{Experimental section}
	\subsection{Graphene ink formulation}
	\begin{figure}[t]
		\centering
		\includegraphics[width=0.90\textwidth]{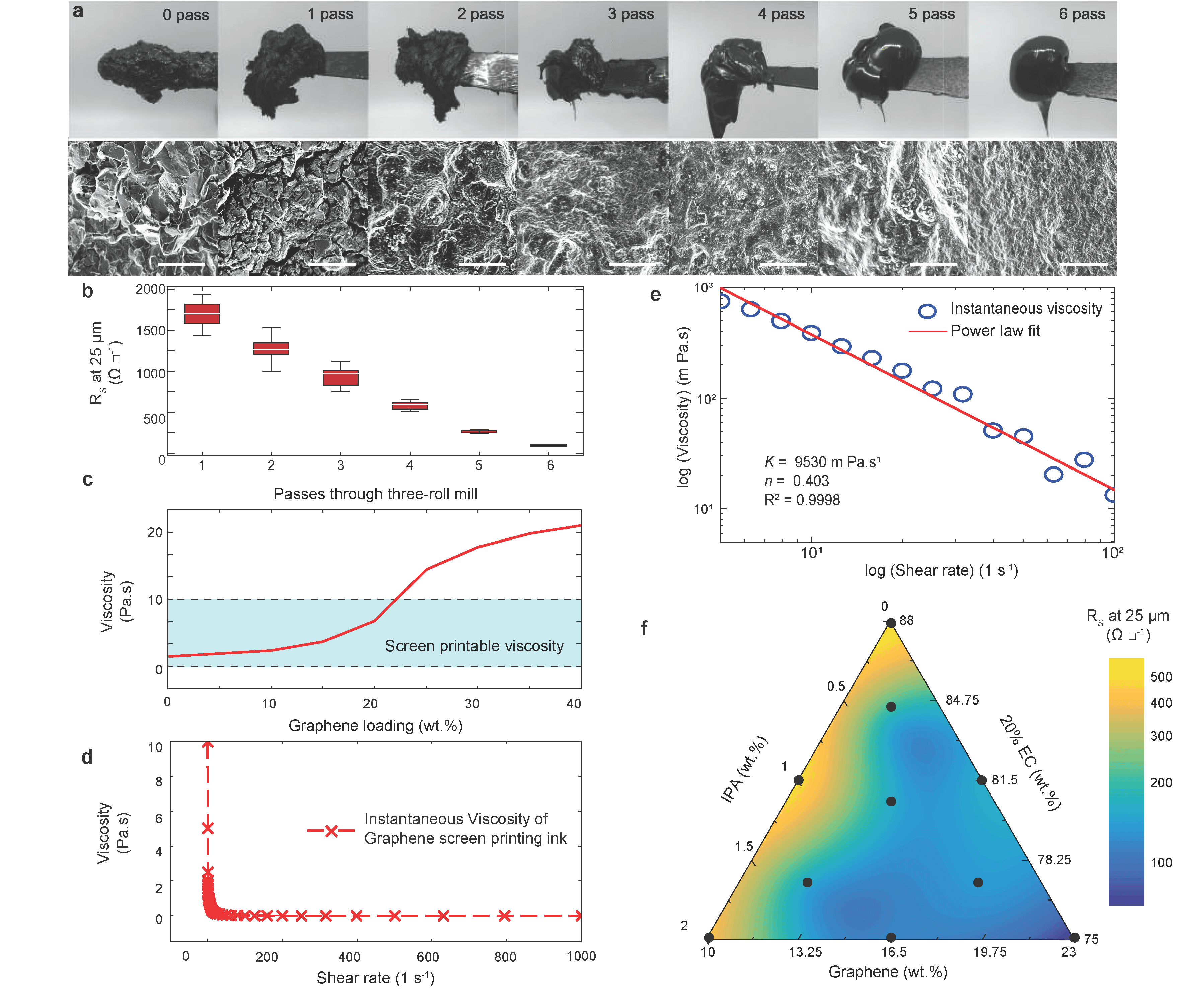}
		\caption{Graphene ink formulation \textbf{a} Photographs of the graphene ink as it is passed through the three-roll mill and corresponding SEM images. Scale bar 5 $\mu$m. \textbf{b} Effect of three-roll milling on the R$_s$ of the ink. \textbf{c} Graph showing the influence of graphene ink on the final ink's viscosity after 6 passes through the three-roll mill. \textbf{d} Rheogram of the graphene screen printable ink, showing shear thinning behaviour. \textbf{e} Log-log plot of viscosity and shear rate of the  power law region of a flow curve. \textbf{f} Ternary plot of the interaction of different loadings of IPA, EC binder and graphene as determined by DOE. Experiments are indicated with black dots. }
		\label{panel1}
	\end{figure}
	
	Like the majority of the graphics inks~\cite{R.H.Leach1993, Goldschmidt2003, Flick1999}, the main components of the functional graphene ink is typically made up of a functional material (in this case, graphene), a binder system (a polymeric film former to provide mechanical integrity after curing) and appropriate solvents (ink carrier).
	Due to the need for the graphene ink to be insoluble during the water-assisted conformal printing, a water insoluble binder, ethyl cellulose (EC), is chosen.
	Terpineol is chosen as the primary solvent due to its compatibility with EC. 
	The high boiling point of the primary solvent terpineol ($\sim$218 $^{\circ}$ C) allows for a long `dwell time' on the screen.
	This prevents clogging of the screens during or immediately after printing\cite{Ng2019a}.
	Isopropyl alcohol (IPA) (boiling point $\sim82.3 ^{\circ}$ C), is chosen as the secondary solvent for on press ink viscosity control.
	As the inks are printed using a hand-operated press, the inks do not have to travel from surface to surface within a printer and is hence subject to less agitation. 
	This in turn prevents common printing issues such as foaming (the formation of air bubbles in the ink due to mechanical agitation usually caused by printing machines) and orange peeling (a mottle-like print defect characterised by dimples in the finished print, giving the print the texture of an orange peel. Usually caused by a failure of the ink to flow out in a smooth, uniformly thick film) \cite{Goldschmidt2003}.
	Therefore, an additive package (i.e. defoamers, surface tension modifiers) that are commonly used in commercial formulations is not included in our ink formulation.
	
	The high viscosity of our ink means that a suitable processing method is required to sufficiently disperse the agglomerates and aggregates of the graphene flakes into primary particles \cite{Ng2019e}.
	Hence, three-roll milling is selected as the main method of dispersion. 
	A three-roll mill consists of a feed roll, a center roll and an apron roll that rotate in opposing directions and at different speeds.
	It is commonly used in ink, paint and coating industries to disperse pigment particles within high viscosity liquids \cite{Goldschmidt2003}. 
	As a dispersion method, it utilises the internal shear and impact forces generated from the movement of the three rollers to break down agglomerates and aggregates into primary particles.
	
	In our formulation, the EC is first mixed with graphene powder (CamGraph G3, Cambridge Nanosystems, avg. flake thickness $\sim$12.4 nm, avg. flake size $\sim$94.5 nm) until the premix becomes a coarse slurry; Fig. \ref{panel1}a.
	We note that the dimensions of the graphene powder fall out of the scope of nomenclature postulated by academic literatures which define 2D graphitic materials with a thickness and/or lateral dimension less than 100 nm as ``Graphite nanoplates'', ``graphite nanosheets'' or ``graphite nano-flakes'' \cite{Bianco2013}. This is to distinguish from finely milled graphite powders whose thickness is typically $>$100 nm and Few Layer Graphene (FLG) with layer numbers of 2 to about 5 and Multi-Layer Graphene (MLG)  with layer numbers of between 2 and 10. We use the term ``graphene powder'' to describe the starting 2D graphitic material as it is a term used by the manufacturer.
	The slurry is then processed by the three-roll mill six times until it assumes a glossy, paste-like appearance.
	Figure \ref{panel1} depicts photographs and scanning electron microscope (SEM) images of the graphene ink as it is processed through the three-roll mill from pass 1 to 6.
	With subsequent passes, improved dispersion of the graphene flakes into primary particles within the ink gradually lowers the sheet resistance (R$_s$) and measurement variability from different areas of the print; Fig. \ref{panel1}b.
	The R$_s$ saturates at 6 passes and does not improve further with increasing three-roll milling; Supplementary Fig. 6. 
	In our experiments, it is able to achieve separation of the graphene agglomerates from an average flake diameter of $\sim$90.5 nm  to $\sim$40.6 nm and an average thickness from 12.4 nm to 2.5 nm; supplementary Fig. 2. 
	Raman measurements of the graphene powders before and after ink formulation do not show any noticeable changes; supplementary Fig 3. 
	
	We observe no major variations and shifts in either the $G$ peak, $2D$ peak or I($D$)/I($G$) ratio indicating that three-roll milling during the ink formulation does not cause any significant morphological or chemical change of graphene. On the other hand, AFM data shows a reduction in both the flake lateral dimension and thickness despite having no observed changes in the Raman spectrum; supplementary Fig. 2. Hence the combination of AFM data and Raman data indicates three-roll milling provides a good method of mechanically dispersing the aggregates of the starting graphene powder within the EC polymeric binder to improve percolation.

	In considering the screen printing of the water-insoluble graphene ink, the formulation must be designed to be homogenized (well dispersed graphene flakes), stable (against aggregation or precipitation) and have thixotropic (shear thinning) rheological properties \cite{Ng2019e}. 
	This is shown by previous demonstrations such as those in references \cite{Arapov2016} and \cite{Karagiannidis2016}.
	As the ink is drawn along the screen \emph{via} the squeegee during printing, the shear stress rate
	($\dot{\gamma}$) acting upon the ink changes from very low ($\sim$1$ s^{-1}$) to very high ($\sim$1000$ s^{-1}$) \cite{Ng2019a, Dybowska-Sarapuk2015} while the viscosity ($\eta$) decreases \cite{Ng2019a, Rosu1999}.
	Upon deposition, $\dot{\gamma}$ decreases as shear rate from the squeegee is removed while $\eta$ increases to form a solid, unbroken image on the substrate surface with a rectangular cross sectional profile. 
	The narrowest printed lines we have achieved on PET are of 200$\mu$m width with a line edge roughness of 4.04 $\mu$ m.; Supplementary Fig. 5.
	To ensure our formulated ink has these properties, a combination of OFAT, for qualitative screening and DOE for quantitative optimisation are utilised. A detailed explanation on the use of OFAT and DOE is given in supplementary note 1. 
	
	We use OFAT to firstly optimise the rheology of the ink by varying the graphene content, thereby determining the maximum loading of graphene permissible in the EC binder before the ink becomes unprintable. 
	OFAT is a method of experimentation that varies one factor at a time in order to find an optimum response and is highly useful for bi-variate analysis \cite{Ng2019e, Jacquez1998}.
	We define the screen printability of the ink as having starting viscosity under zero shear being in between 2 to 10 Pa s \cite{Ng2019e, Hu2018}. 
	To derive the maximum loading of graphene, first 20 wt.\% of EC powder is fully dissolved in terpineol \emph{via} stator-rotor mill, creating the EC binder. 
	The graphene powder is then dispersed in this EC binder \emph{via} three-roll mill to ascertain the maximum loading of graphene before the ink becomes unprintable. 
	We find that the maximum amount of graphene that can be introduced to the EC binder before the viscosity of the ink exceeds the upper limit of the screen printable viscosity range ($\sim$2-10 Pa s) is $\sim$23 wt.\%; Fig. \ref{panel1}c. 
	In addition, rheological characterisation is also carried out on the ink at this stage to ensure it has the necessary shear-thinning properties as depicted in \ref{panel1}d for screen printing. 
	To further characterise shear thinning behaviour, the measured viscosity and shear rate were plotted in a log-log plot to observe power law behaviour; Fig. \ref{panel1}e. 
	The power law model, also known as the Ostwald de Waele model is a rheological model that can be applied to shear thinning fluids to observe their behaviour under shear \cite{Bjrn2012}.
	The model can be expressed mathematically using the following equation:
	
	\begin{equation}
	\eta = K  ^* \dot{\gamma}^{n-1}
	\label{powerlaw}
	\end{equation}
	
	where $K$ is the consistency factor and $n$ is the power law index; $n <$1 represents shear thinning behaviour, $n$ = 1 represents Newtonian behaviour and $n >$1 = shear thickening behaviour \cite{Bjrn2012}.
	The resultant curve with a power law model, yield a good fit to the flow curve (R$^2$ = 0.9998) between the values 5 to 10 $^2$ s$^1$ s$^{-1}$.
	We apply a curve fitting based on equation \ref{powerlaw} to the log-log plot which determine $K$= 9530 m Pa.s$^n$ and $n$ = 0.403, indicating that the graphene ink behaves as a viscous fluid that is very shear thinning.
	
	Using the results of the first set of OFAT experiments, we establish a realistic range of parameter levels, based on the three basic ink components- graphene (10 – 23 wt.\%), EC binder (75 – 88 wt.\%) and IPA (0 – 2 wt.\%) as a solvent. 
	These ranges are used as the upper and lower limits for the DOE to optimise the conductivity of the ink. 
	We note that the above ranges may differ should graphene from a different source be used.
	DOE is a widely-used statistical approach to experimental formulation in screening and optimising parameters (also known as factors) \cite{Ng2019e, Jacquez1998} and multivariate analysis.
	Inks are produced according to the parameter ranges indicated in Fig. \ref{panel1}e and plotted in a manner reflecting the interaction between graphene, IPA and EC binder.
	The experiments and results of the DOE are given in supplementary information and have also been plotted in Fig. \ref{panel1}f in a ternary plot reflecting the interaction between the three ink components. 
	An optimum formulation for the screen printing graphene ink is established as graphene (23 wt.\%), EC binder (75 wt.\%) and IPA (2 wt.\%) with a R$_s$ response of 64 $\Omega$ $\square^{-1}$ while on PET.
	The formulated ink is stable and not prone to sedimentation for more than 1 month; Supplementary Fig. 4.
	The lowest measured value of R$_s$ at 25 $\mu$m is 24 $\Omega$ $\square^{-1}$ when deposited on glass (bulk conductivity = 0.17 $\times$ 10$^4$ S m $^{-1}$).
	We stress that this value represents the bulk conductivity of the screen printed ink without further annealing, compression, and partial or full decomposition of the non-conductive polymeric binder.
	As will be discussed later, this variation in R$_s$ is attributed to differences in surface roughness.
	We further note that the above process optimises the electrical conductivity while maintaining the ink printability, and to our knowledge, have not been methodically applied to formulation of graphene or other 2D material inks in published literature.
	Although the final conductivity value will vary depending on the source of graphene and type of binder used, the above discussion provides a systematic and universal approach to achieve optimised ink formulation.

	\subsection{Water-assisted conformal printing method}
	
	\begin{figure}[!t]
		\centering
		\includegraphics[width=\textwidth]{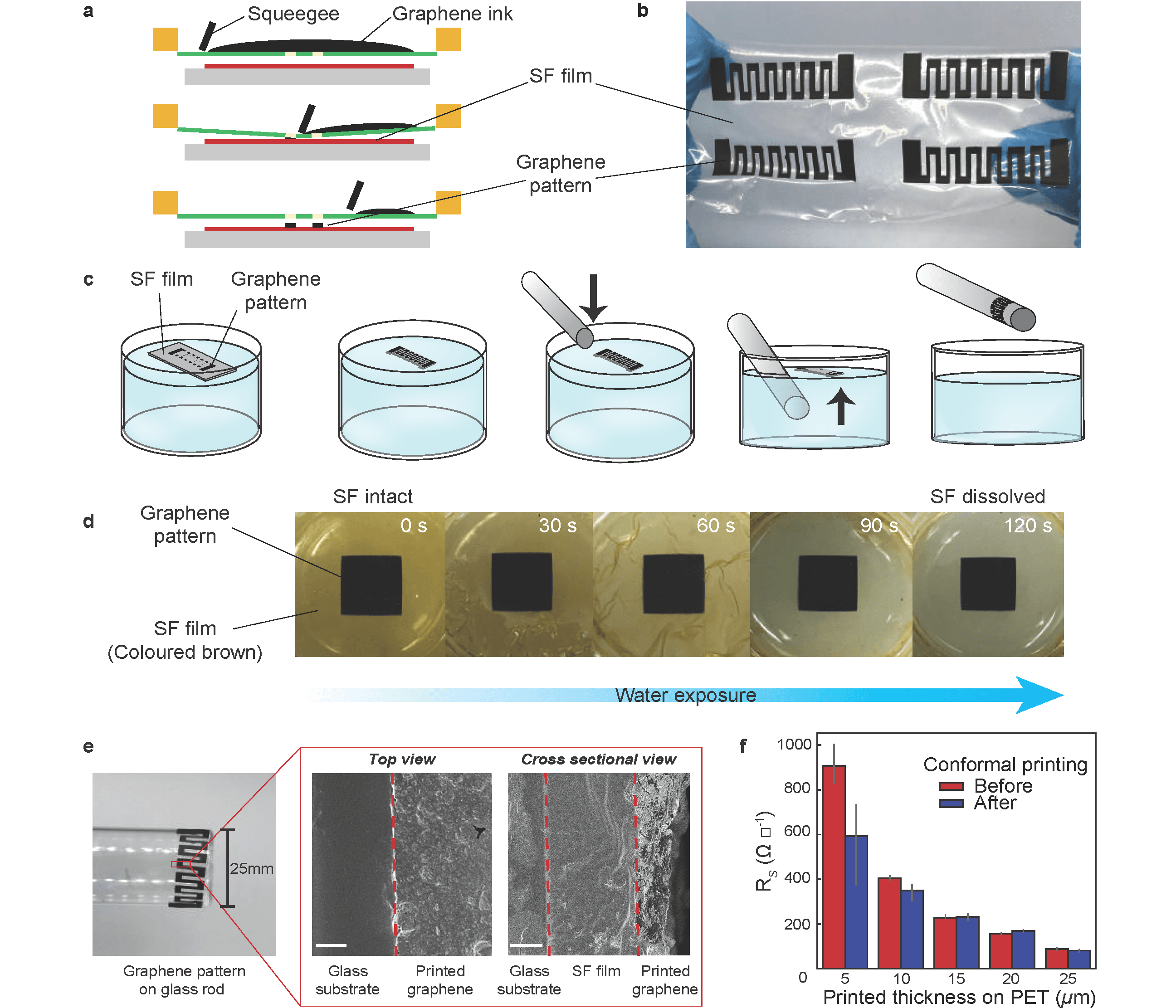}
		\caption{Water assisted conformal printing. \textbf{a} Schematic of screen printing of graphene ink on to the SF. \textbf{b} Screen printed conductive pattern array on water-soluble SF. \textbf{c} Schematic of the water assisted conformal printing process starting with the floating of the printed SF on water followed by the  dissolution of the surrounding SF, leaving behind residual wet SF underneath to act as an adhesive to the substrate. Immersion of the substrate into the water bath to `fish' the conductive pattern out which eventually adheres to the substrate by drying of the SF. 
			\textbf{d} Interaction of the printed graphene ink with water soluble SF over 120 s. The SF is coloured using a food dye.
			\textbf{e} Photograph of the graphene pattern on a glass rod with representative SEM views of the top and cross section. Scale bar 5 $\mu$m \textbf{f} Comparison of R$_s$ before and after water assisted conformal printing across a variety of printed graphene ink thicknesses on PET.}
		\label{panel2}
	\end{figure}
	
	Figure \ref{panel2}a-g schematically illustrates the key steps in the water-assisted conformal printing process.
	This involves firstly screen printing the formulated graphene ink directly on to a thin ($\sim$30 $\mu$m), water soluble SF of PVA that has been previously cast on a polytetrafluoroethylene (PTFE) plate using screen printing; Fig. \ref{panel2}a. 
	For this, the water-insoluble graphene ink is first flooded over the screen.
	A squeegee is then drawn across it as shown in figure \ref{panel2}a, applying shear forces to push the ink through the open pores of the screen \cite{Ng2019a} directly on to the SF, producing an array of conductive patterns (length, 200 mm; height, 50 mm per pattern); Fig. \ref{panel2}b.
	When printing multi-layered structures, the first deposited layer of ink is allowed to dry before the next layer is printed over the first print.
	The SF containing the single and multi-layered patterns are then detached from the PTFE plate \emph{via} simple peeling.
	
	The printed patterns on the SF are then floated on to the surface of a de-ionised water bath (25 $^{\circ}$C); Fig. \ref{panel2}c.
	This allows the dissolution of PVA ($\sim$120 s PVA dissolution time), while retaining the single and multi-layered patterns which are water insoluble with the exception of an ultrathin layer of still wet PVA SF ($\sim$20 $\mu$m); Fig. \ref{panel2}d.
	The object to be printed on is then immersed into the water bath to `fish' out the conductive pattern, allowing the pattern to adhere to it with the wet SF acting as the main interface between the ink patterns and the substrate; Fig. \ref{panel2}e, f. 
	The printed pattern conforms directly to the topography of the 3D object. 
	Upon desiccation, the residual SF under the conductive patterns dries, reduces in thickness to $\sim$15 $\mu$m and acts as an adhesive for the graphene patterns directly on to the substrate; Fig. \ref{panel2}e.
We note that this process is still in the proof-of-concept stage and will require further optimisation before it can be scaled-up.
	
	The success of this method of conformal printing relies on the interaction between the water-insoluble graphene ink previously printed in a pattern with the water soluble SF film. 
	To illustrate this in figure \ref{panel2}d, the water-soluble PVA is artificially coloured brown with food dye and a water-insoluble graphene pattern printed on to it \emph{via} screen printing. 
	This composite structure is then exposed to water and observed for 120 s which is the time needed for the water-soluble SF to fully dissolve around the printed pattern at 25 $^{\circ}$ C, while leaving the graphene pattern fully intact (Fig. \ref{panel2}d), with the ultrathin PVA layer underneath acting as the ink-substrate interface.

	Figure \ref{panel2}e shows the final printed pattern on a glass rod and corresponding SEM images of the top and cross-sectional view of the printed conductive pattern, revealing the dried SF film ($\sim$15 $\mu$m) acting as the adhesive layer between the conductive pattern and the glass substrate.
We also carry out a pull off adhesion test of a transferred pattern on a glass substrate using a BGD 500 adhesion pull off tester. 
From three separate measurements, we find that it requires an average of 8.8 $\times$ 10${^5}$ N m${^{-2}}$ of force to detach the pattern from the substrate.
   We observe a $\sim$40 \% improvement in adhesion over time as the same set of three adhesion tests carried out three months later on the same sample requires an average of 1.25 $\times$ 10${^6}$ N m${^{-2}}$ of force to pull off.
   This is comparable to our measurements on commercial carbon inks printed on glass substrates (1.1 $\times$ 10${^6}$ N m${^{-2}}$).
	Representative sheet resistance (R$_s$) data recorded for the patterns show no significant increase in R$_s$ before and after transfer; Fig. \ref{panel2}f. 
	At thinner prints of graphene ink ($\sim$4 $\mu$m), there is an observed improvement in R$_s$, presumably due to polymer shrinkage after the dissolution of the SF. 
	As the printed graphene ink thickness increases to 25 $\mu$m, the electrical response becomes increasingly monotonic as R$_s$ drops to 71$\sim$88 $\Omega$ $\square^{-1}$ after transfer.
	We note that this is achieved without any post-processing other than drying at ambient temperature ($\sim$25 $^{\circ}$C).
	The predictable change in R$_s$ after the transfer process underscores the potential of this technique to fabricate electrical devices with well-defined performance parameters on to arbitrarily-shaped objects without the need for any post deposition treatment.

	\begin{figure}[t]
		\centering
		\includegraphics[width=0.85\textwidth]{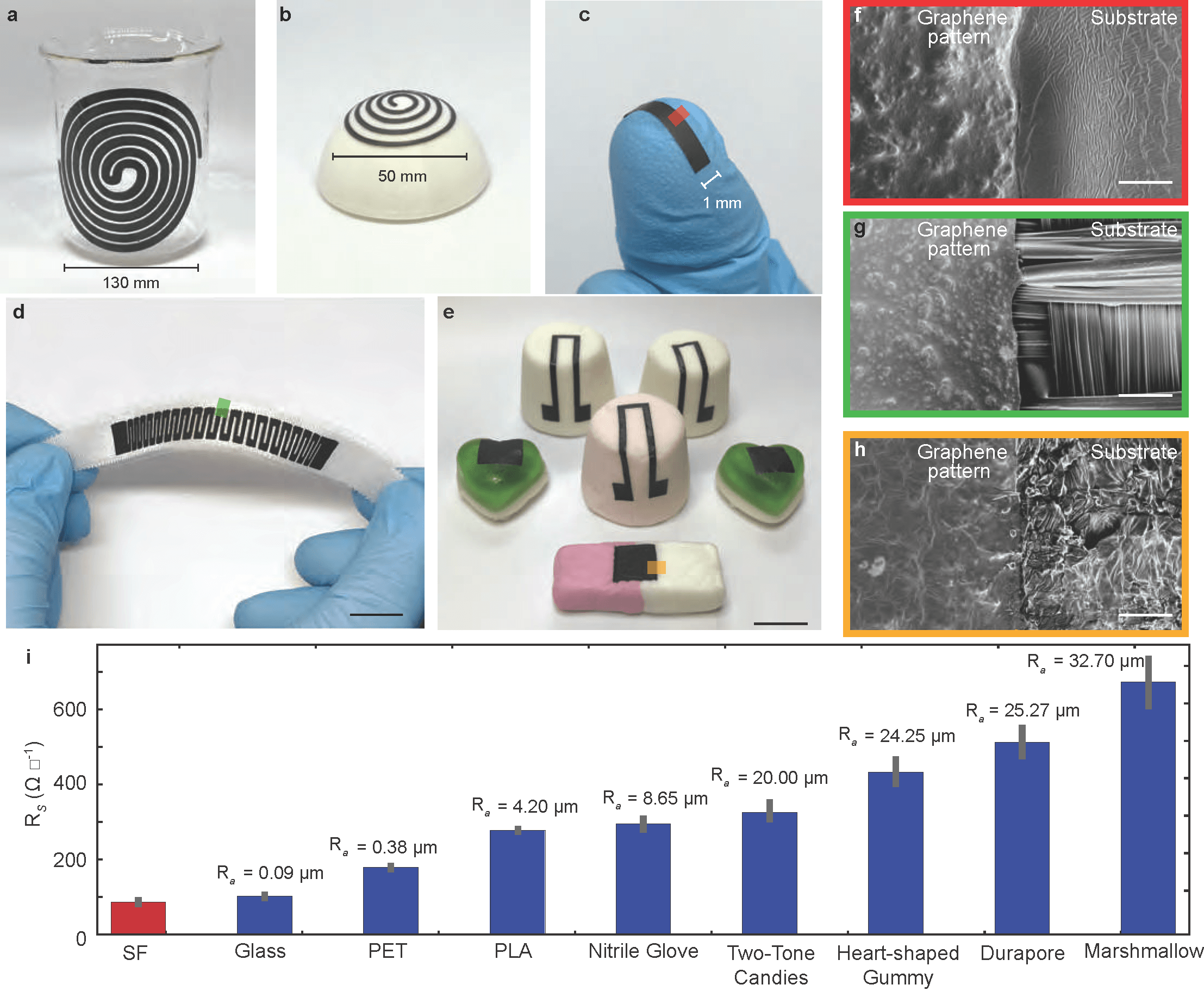}
		\caption{Conformal printing on to various substrates. \textbf{a} In a spiral pattern on glassware. \textbf{b} On a 3D printed thermoplastic dome. \textbf{c} A 2 mm graphene strip on to a latex glove on a finger. \textbf{d} On a 3m textile bandage in a sensor pattern. \textbf{e} On to pink and white marshmallows, green heart-shaped gummy candies and two-toned white/pink candies. Scale bar 1 cm.  Top view SEM images of the \textbf{f} section highlighted in red in \textbf{c}, \textbf{g} section highlighted in green in \textbf{d} and \textbf{h} section highlighted in orange in \textbf{e}. Scale bar 5 $\mu$m. \textbf{i} R$_s$ of patterns after water assisted printing on different substrates correlated with R$_a$.}
		\label{panel3}
	\end{figure}
	
	To highlight the versatility of this technique, we print single and multi-layered patterns on to various arbitrarily shaped objects: (1) hard, conformal, (2) flexible and (3) soft, conformal substrates, even with sharp angles; Supplementary Fig. 8.
We note that the process is most proficient when printing on surfaces with limited curvature.
		Certain shapes, especially 3D concave surfaces (e.g. the inside of a dome), are difficult to print on using this method.
	In demonstrating the ability to print on hard conformal substrates, we create spiral conductive graphene patterns directly on to the surface of a glass beaker (figure \ref{panel3}a) and on a 3D-printed dome made of polylactic acid (PLA), a thermoplastic; Fig. \ref{panel3}b.
	In both cases, the resulting prints yield continuous lines that are electrically conductive in their entirety without damaging the original dimensions of the pattern (R$_s$ $\sim$120 $\pm$10 $\Omega$ $\square^{-1}$ at $\sim$25 $\mu$m before and after transfer).
	Overall, the process is suitable for most hard surfaces and there is no significant change in the dimensions of the printed lines at $\sim$25 $\mu$m thickness or structures before and after printing.
	Large-device prints (130 mm $\times$ 130 mm) can also be achieved, especially on the glass beaker.
	The resolution of the deposited pattern (i.e. edge definition) is also highly dependant on the printing method employed in the process.
	We therefore propose that this water-assisted conformal printing process will work just as well with other high-throughput forms of printing, including flexography, rotogravure and inkjet printing \cite{Ng2019a, R.H.Leach1993}.
	We also note that thin lines like those in Fig. \ref{panel3}b run the risk of deformation during the dissolution phase due to the natural, uneven dissolution of the SF film in water at 25 $^{\circ}$C. 
	As such, the prints in Fig. \ref{panel3}(a, b) are reliably achieved only with a precise 120 s dissolution time in our experimental setup.
	
	Adhesive bandages are used daily to protect small wounds and adhere securely to the body and conform to every contortion of the skin.
	Similarly, nitrile gloves adapt and conform well to the human body.
	To date, research has gone into developing sensors that are fabricated on to bandages \cite{Yamada2011} and developing other forms of human motion detection devices on to gloves \cite{Zhao2017} for wearable applications.
	To demonstrate the ability of this strategy to conformally print on to flexible substrates, we therefore use nitrile gloves and bandages as target substrates. 
	Using the method described above, a thin 1 mm width strip is conformally printed directly on the tip of the finger of a latex glove.
	A serpentine pattern which could potentially be used for strain sensing is also conformally printed on to an off-the-shelf 3M Durapore\textsuperscript{TM} textile bandage. 
	SEM images at the interface of the conductive graphene patterns and the substrates (nitrile glove: Fig. \ref{panel3}f, Durapore bandage: Fig. \ref{panel3}g, candies: Fig. \ref{panel3}h) reveal the nature of the water transfer process whereby the functional patterns are adhered and `sit' atop the substrates as compared to being printed directly.
We note that the adherence of the printed patterns to the substrate is highly dependent on a water-soluble PVA layer which renders the final product water-sensitive. Therefore the potential of the final application being exposed to water must be considered.
Additionally, we recognise that many functional devices may include water-sensitive materials and their compatibility with this method must also be taken into account.
	
	Finally, we evaluate the ability of the water-assisted conformal printing method to print on soft substrates. 
	To demonstrate this, we print a series of patterns on assorted candies including marshmallows, green heart-shaped gummy candies and two-toned white/pink candies; Fig. \ref{panel3}e. 
	Despite mild deformation, the patterns are able to maintain their shape, electrical conductivity and adhesion to their respective substrates.
	Printing conductive patterns directly on to conventionally non-functionalised surfaces such as those for candies allows for the fabrication of small form factor devices which will be further demonstrated in the following section.
	
	We note that the properties of the water-assisted conformally printed structures may be affected by the roughness of the target substrate. 
	We take average roughness (R$_a$) measurements \emph{via} profilometry of the various above-mentioned substrates and find that increased R$_a$ is correlated with increased R$_s$ of single-layer water-assisted conformal patterns; Fig. \ref{panel3}i.
	This is attributed to the printed patterns taking on the different unique morphologies of the respective substrates.
	The process we present is a proof-of-concept demonstration of conformal printing. 
One particular area for further work is in effective registration of the functional patterns.
To achieve this, two strategies that are compatible with our process and could potentially be used are the water-level lowering method \cite{Saada2017} or automatic registration using robotics \cite{LeBorgne2017}.
	
	\subsection{Applications of water-assisted conformal printing}
	
	We have already shown that water-based conformal printing can be used as a simple and effective method to create functional printed patterns on to different substrates while maintaining the properties of the structure.
	This allows fabrication of functional devices directly on to previously-inert surfaces. 
	To demonstrate this, we conformally-print devices, including joule heaters on glassware, conductive patterns on small gummy candies for capacitive based proximity sensing and deformation sensors on marshmallows and candies.
	We also demonstrate the ability of this method to conformally print multi-layered devices by demonstrating the fabrication of a graphene/PU/graphene thin film parallel plate capacitor and subsequent water transfer on to soft and hard substrates. 
	
	\begin{figure}[ht!]
		\centering
		\includegraphics[width=0.8\textwidth]{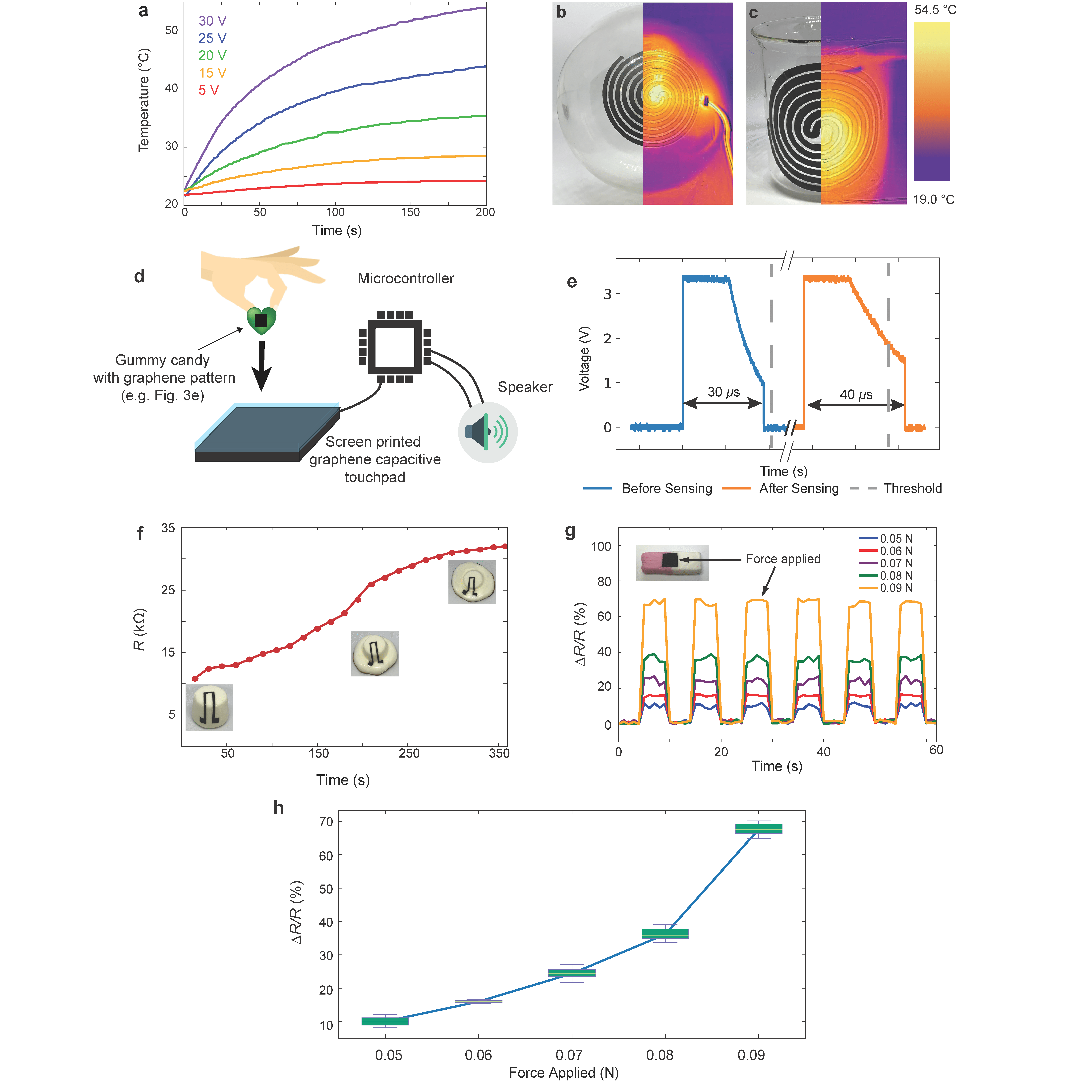}
		\caption{Applications of water assisted conformal printing. \textbf{a} Temperature evolution curves of conformally printed spiral heater under constant voltages of 10 V, 15 V, 20 V, 25 V and 30 V. \textbf{b} Photograph of a spiral pattern functioning as a joule heater on a round bottom flask with the corresponding IR image. \textbf{c} Photograph of a spiral pattern on a glass beaker with a corresponding IR image. \textbf{d} Conformally printed square pattern functionalising green heart-shaped candies for capacitive touch sensing. \textbf{e} Circuit discharge time before and after proximity capacitive sensing of the candy. \textbf{f} Conductivity response of the deformation of a marshmallow’s surface. \textbf{g} Deformation sensor printed on two-toned white/pink candies with corresponding $\Delta R/R$ responses when applied with increasing levels of force. \textbf{h} Average $\Delta R/R$ response in relation to increasing levels of force.}
		\label{panel4}
	\end{figure}
	
	Joule heaters fabricated \emph{via} printing nanomaterials \cite{Kim2013} such as graphene \cite{Kang2011,Sui2011} are of great research interest. 
	Joule heating, also known as ohmic heating is the process by which the passage of an electric current through a conductor produces heat and is currently the primary method by which many thin-film heaters function.  
	Printed graphene heaters are have huge potential for disposable devices such as wearable heaters for therapeutic purposes \cite{Jang2017}.
	Demonstrations to date on flat surfaces has seen screen printing \cite{He2017} emerging as an attractive method for the large-scale fabrication of joule heaters.
	This is due to its ability to print thicker layers of functional materials accurately compared to other conventional printing techniques.
	We demonstrate the potential of our water-assisted conformal printing method to fabricate joule heaters on arbitrary substrates.
	
	To realise joule heaters, we conformally print an unbroken spiral pattern (130 mm $\times$ 130 mm) on to a beaker and a round bottom flask; Fig. \ref{panel3}a, b. The prints are allowed to dry in ambient temperature ($\sim$25 $^{\circ}$C).
	We then apply varying voltages of 10 V, 15 V, 25 V and 30 V using a DC power supply producing a maximum of 54.5 $^{\circ}$C of heating in 200 s; Fig. \ref{panel4}a. 
	Representative IR images of heating are depicted in Fig. \ref{panel4}(b, c).
	The heater functions consistently throughout all voltage inputs, demonstrating how water-assisted conformal printing could be used to create small devices on arbitrarily shaped, previously non-conductive surfaces. 
	
	We next demonstrate the use of water-assisted conformal printing for proximity capacitive sensing. 
	A proximity capacitive sensing device consists of two conductive electrodes separated by a dielectric. 
	To realise this, we first conformally print a conductive graphene pattern on a candy (\emph{T} electrode). 
	The other conductive graphene pattern (\emph{S} electrode)is screen printed on to the reverse side of PET where PET acts as the dielectric. 
	We use a 16-bit microcontroller that continuously charges the \emph{S} electrode every 250 ms intervals to 3.3 V, generating a surrounding electric field. 
	After a fixed time period at 3.3 V, the microcontroller starts to discharge the \emph{S} electrode through a 512 k\textit{$\Omega$} resistor.
	This voltage is brought to 0 V through a Schmitt trigger if the \emph{S} electrode discharges to below 1 V or if the overall discharge time exceeds 40 $\mu$s.
	A threshold of 35 $\mu$s (denoted by a vertical dashed line; Fig. \ref{panel4}e) is set to trigger a response from the microcontroller.  
	As seen in Fig. \ref{panel4}e, the discharge time for the \emph{S} electrode alone (i.e. before sensing) is $\sim$ 30 $\mu$s.
	When the \emph{T} electrode is brought in close proximity to the \emph{S} electrode, it causes a disruption in the electric field, resulting in an increase in overall capacitance. \cite{Yao2014,Hayashi2005,Cotton2009}.
	This is manifested by an increase in the RC constant, and therefore, discharge time of the \emph{S} electrode beyond the set threshold of 35 $\mu$s; Fig. \ref{panel4}e, triggering a response from a speaker connected to the microcontroller. 
	The sensing process is shown in the supplementary video.
	This demonstrates how conformally printed conductive patterns could enable any object for proximity capacitive sensing in a simple, cost-effective manner. 
	
	\begin{figure}[t!]
		\centering
		\includegraphics[width=0.9\textwidth]{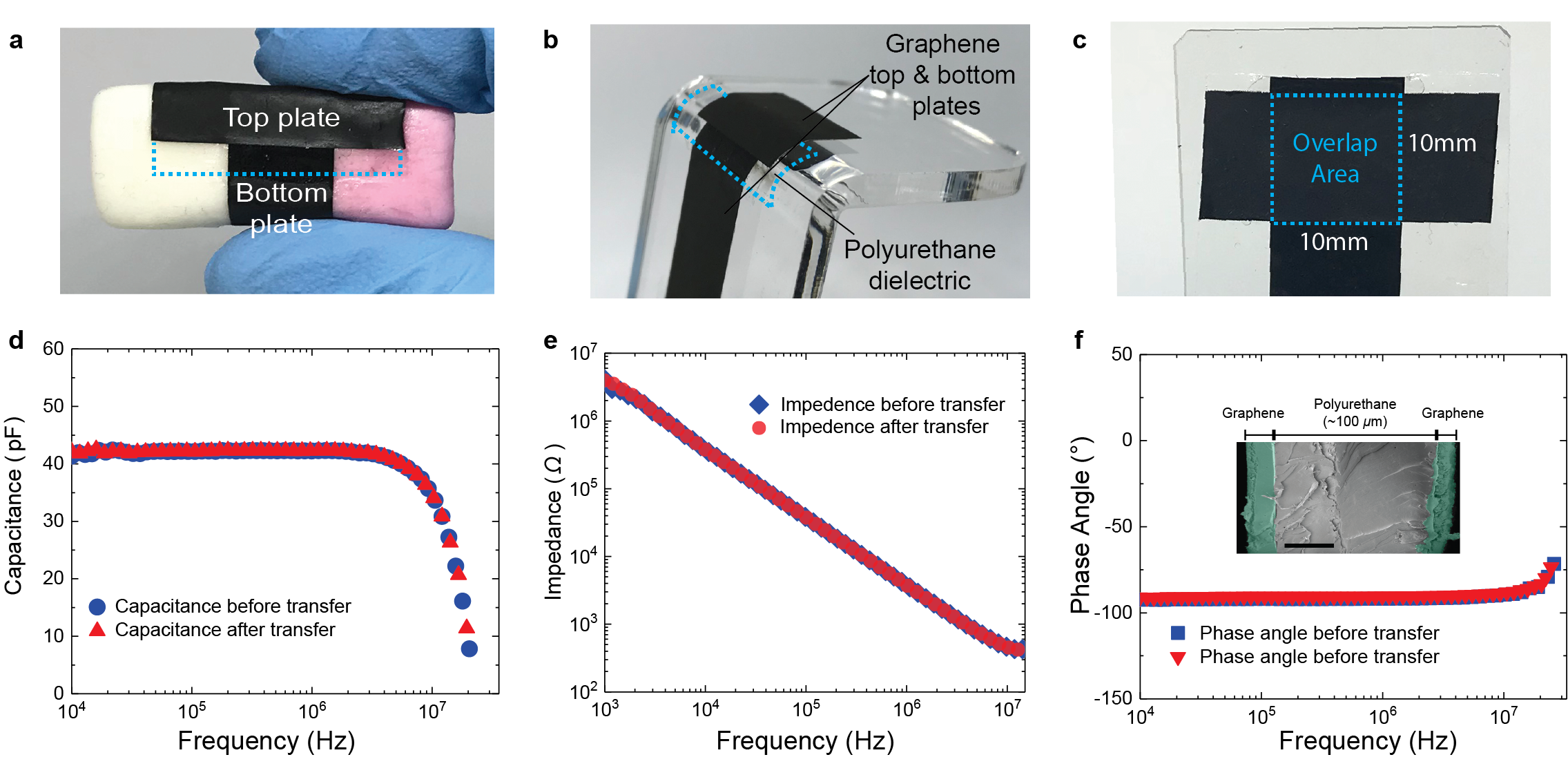} 
		\caption{Conformal printing of parallel plate capacitors. Photographs of a capacitor on \textbf{a} white/pink candy, \textbf{b} acrylic and \textbf{c} glass substrate. 
			Frequency response of \textbf{d} the capacitance \textbf{e} Impedance value and \textbf{f} phase angle of the capacitor before and after conformal printing. Inset: SEM image of cross-section of printed graphene/pu/graphene parallel plate capacitor. The graphene layers have been false-coloured for clarity. Scale bar: 30 $\mu$m}
		\label{panel5}
	\end{figure}
	
	Graphene based sensors capable of detecting motion \emph{via} resistive changes responding to force have already been widely researched \cite{Amjadi2016, Tian2014} and have been used in a variety of applications including human-motion detection.
	We demonstrate the versatility of our printing method in the fabrication of sensors functioning in a similar manner capable of detecting deformation on soft substrates.
	For this, we conformally print a graphene pattern directly on to the surface of a marshmallow to measure its surface deformation when heated at a constant temperature.
	The marshmallow is placed on a hotplate pre-heated to 300 $^{\circ}$C and resistance readings of the deformation sensor are taken every 5 s for a period of 360 s; Fig. \ref{panel4}f(inset).
	The melt profile allows us to monitor the corresponding structural deformation in real time.
	In a second resistive deformation sensor demonstration, we conformally print a conductive square of graphene ink directly on a white/pink candy; Fig. \ref{panel4}h.
	Increasing levels of force of 0.05 N, 0.06 N, 0.07 N, 0.08 N and 0.09 N are applied at 5 second intervals to the square of graphene as shown in Fig. \ref{panel4}g. 
	Two strips of copper tape are attached to the edges of the printed graphene pattern as terminals and the resistance measured using a multimeter as force is applied.    
	This surface deformation sensor can achieve a relative change in resistance ($\Delta R/R$) upon application of force in a repeatable fashion due to the elasticity of the substrate. 
	The $\Delta R/R$ response for the different forces applied is linear up 0.08 N; see Fig. \ref{panel4}h. 
	At 0.08 N of force, the trend appears to become non-linear, indicating that the graphene pattern is approaching its elastic limit. 
	Additional force applied past 0.09 N no longer generates an effective resistive response as resistances do not return to baseline levels ($\Delta R/R$ = 0 \%). 
	This indicates that the possible effective range of the deformation sensor on this substrate is between 0.05 N to 0.08 N.
	The ability to conformally print directly on to small, soft and delicate surfaces again highlights the versatility of water-based conformal printing to produce small item sensors that are able to provide comprehensive information for a wide variety of applications.
	For instance, in our demonstration, the ability to print directly on to food items and derive physical information such as deformation rate could be applied to other fields in food and agriculture.

	The above examples showcase the ability of water-assisted conformal printing in the fabrication of single layered devices only.
	However, the use of conventional printing techniques in tandem with conformal printing also means that this method could be easily adapted to printing multi-layered devices such as parallel-plate capacitors.
	The capacitors fabricated here consist of printed, stacked layers of graphene/PU/graphene.
	The bottom electrode is first deposited \emph{via} screen printing on to the SF.
	This is followed by blade applicator coating of pre-polymerised, liquid PU which is then allowed to cure at room temperature, producing a $\sim$100 $\mu$m thick dielectric layer.
	Finally, we screen-print the top graphene electrode in a manner similar to the deposition of the bottom electrode over the cured PU film, forming a graphene/PU/graphene capacitive structure (10 mm $\times$ 10 mm overlap area). 
	The capacitor is then conformally printed on to a variety of substrates including glass, acrylic and white/pink candies; Fig. \ref{panel5}a-c.
	
	To characterise the properties of the capacitor printed on to the glass substrate, impedance analysis is carried out before and after the transfer process. 
	Figures \ref{panel5}d-f present the frequency response of the capacitor with a 10 mm $\times$ 10 mm electrode overlap area (as shown in Fig. \ref{panel5}c) and a dielectric thickness of $\sim$100 $\mu$m (measured from the cross section SEM image in Fig. \ref{panel5}f: inset). 
	The capacitance ($C$) ($\sim$40pF) and impedance ($|Z|$) across the measured range (10$^4$ $\sim$ 10$^7$ Hz) before and after water-assisted printing does not show any significant change; Fig. \ref{panel5}d, e.
	The data recorded for the phase angle indicates a dominating capacitive component with a corresponding phase angle of -90 $^{\circ}$ from 10$^4$ Hz to 10$^7$ Hz, consistent with the operations of a typical printed capacitor with an R-C equivalent behaviour; Fig. \ref{panel5}f.
	The relative dielectric constant ($\epsilon{_r}$) extracted from the measurements gives a value of $\sim$4.6 through the frequency range 10$^4$ $\sim$ 10$^7$ Hz.
	This is in agreement with previous literature stating the $\epsilon{_r}$ of a pure PU matrix is $\sim$3.2-8.0 \cite{Li2016, Huang2004}. 
	The transfer of the capacitor using waster-assisted conformal printing directly on to arbitrary substrates demonstrates exciting opportunities in using our printing process in the fabrication of more complex multi-layer, multi-component devices and systems from functional inks of 2D crystals. 
	
	\section{Conclusion}
	In this work, we demonstrate a room-temperature method of conformally printing functional structures on an arbitrarily-shaped 3D objects using a water-assisted technique. 
	This is achieved by carefully designing a screen-printable, water insoluble conductive graphene ink \emph{via} OFAT and DOE methods of optimisation. 
	The printing method is highly versatile in its ability to fabricate functional structures on to arbitrarily-shaped surfaces and substrates (glass, 3D printed objects, textiles, nitrile and gummy candies) without the loss of the electrical properties.
	To demonstrate the potential applications of this conformal printing, we fabricate joule heaters directly on to glasswares that are capable of achieving 54.5 $^{\circ}$C in 200 s.
	In another demonstration, we show how small items such as a gummy candy can also be rendered functional to make it capable of proximity capacitive sensing.
	Small-item sensing is also realised in another demonstration where we print versatile and highly sensitive deformation sensors on delicate substrates such as marshmallows and gummy candies.
	Finally, we demonstrate multi-layered device fabrication by conformally printing a graphene/PU/graphene parallel-plate capacitor on both soft and hard substrates without any noticeable change in device parameters.
	Our versatile conformal printing method gives the ability to print devices on different arbitrarily-shaped 3D objects without any harsh chemical or temperature treatment, lending new functionalities to previously-inert surfaces.

	\section*{Methods}
	
	\subsection*{Ink preparation, printing and characterisation}
	The ink binder is first prepared by mixing EC (Sigma Aldrich, 200646, viscosity 4 cP, 5\% in toluene/ethanol) with terpineol (Sigma Aldrich, W304506, $\alpha$-Terpineol) into a glass jar at a concentration of EC:terpineol 20:80 wt.\%.
	The mixture is then mixed in a Silverson L5m laboratory high shear stator-rotor mill at 4,000 rpm for 45 minutes.
	The glass jar sat in a water-bath during dissolution and homogenisation to dissipate the heat generated from the mixing process.
	Commercially-sourced graphene powder (Cambridge Nanosystems) and IPA is then added to the homogenised binder solution in a 23:2:75 wt.\% graphene:IPA:EC and mixed until a coarse slurry is formed.
	The slurry is then put through a three-roll mill (EXAKT 50I) 6 times with roller gaps set to $\sim$30 $\mu$m until a glossy appearance is achieved.
	Prints are carried out using a hand operated screen printing press (80 mesh/inch).
	R$_s$ measurements are taken using a four-point probe system.
	A parallel plate rotational rheometer (DHR rheometer, TA instruments) is used to evaluate the viscosity of the inks as a function of the shear rate.

	\subsection*{SF preparation and characterisation}
	For water soluble SF, PVA (MP Biomedicals, MW 15,000 Da) is mixed with de-ionised (DI) water into a sealed container at a ratio of 20:80 wt.\% and stirred with a magnetic stir bar for 12 h at 400 rpm under 100 $^{\circ}$C.
	To cast the SF film, the obtained solution is bar coated at a 50 $\mu$m wet thickness on to a PTFE plate and is allowed to dry at ambient temperature ($\sim$25 $^{\circ}$C).
	The dry thickness thickness is measured \emph{via} SEM.
	After screen printing and curing of the single or multi-layered functional structures on to the SF, the resultant composite stack is detached from the PTFE plate \emph{via} simple peeling.
	For exposure studies, graphene ink is printed in a square pattern on to the SF colored with food dye, peeled and exposed to water using an enclosed glass observation jar and timed.
	
	\subsection*{Joule heater}
	IR imaging is carried out using a FLIR ONE thermal imaging camera. 
	Temperature curve measurements are carried out using an arduino UNO unit with a TMP-36 temperature sensor attached directly on to the graphene heating element. Readings are sampled every 1 ms and averaged over 1 s.
	
	\subsection*{Proximity sensor}
	The graphene $S$ electrode is fabricated \emph{via} screen printing of the ink in to a single strip and connected to the microcontroller using copper tape.
	Water-assisted conformal printing is used to print 1 cm $\times$ 1 cm squares ($T$ electrode) of graphene directly on to the gummy candies. 
	
	\subsection*{Deformation sensor on marshmallow and white and pink candy}
	A copper tape is attached to the edges of the printed graphene sensors to be used as contacts.
	Resistance measurements are carried out using a Keithley 2100 Digital Multimeter with probes attached directly to the copper tape \emph{via} crocodile clips.
	
	\subsection*{Parallel plate capacitor}
	Commercial pre-polymerised PU is obtained as a 2-component resin and hardener. 
	The two components are mixed (resin to hardener ratio 10:1) and then degassed in a vacuum jar for 10 minutes before application \emph{via} blade applicator as the dielectric layer. 
	A Newtons4th Impedance Analysis Interface 2 with Newtons4th PSM3750 is used to measure the frequency response of the capacitor. 
	
	\subsection*{Scanning electron microscopy}
	SEM images are taken with a high-resolution LEO 1530 VP Ultra-high performance VP SEM system. 
	The field emission gun is operated at an accelerating voltage of 2.4 keV.
	
	\subsection*{Atomic force microscopy}
	Measurements of graphene flakes are taken with a Bruker Dimension Icon in scan assist mode. Samples are diluted 100 times and then drop cast on to pre-cleaned (isopropanol) Si/SiO$_2$ substrate. 
	
	\subsection*{Profilometry measurements}
	Profilometry measurements of the substrates are taken with a Bruker DektakXT stylus profilomter with a scan length of 4 mm, a resolution of 0.66 $\mu$m, stylus force of 3 mg and a stylus tip radius of 12.5 $\mu$m.
	
	\section*{Acknowledgements}
	We acknowledge support from EP/L016087/1, National Research Fellowship (NRF) of Korea and the Graphene Flagship.

	\bibliographystyle{naturemag}
	
	
	
	
\end{document}